\documentstyle[12pt]{article}

\newcommand{\be}{\begin{equation}}
\newcommand{\ee}{\end{equation}}

\newcommand{\bea}{\begin{eqnarray}}
\newcommand{\eea}{\end{eqnarray}}

\newcommand{\NP}[1]{Nucl. Phys.\ {\bf #1}\ }
\newcommand{\PL}[1]{Phys. Lett.\ {\bf #1}\ }

\newcommand{\PR}[1]{Phys. Rev.\ {\bf #1}\ }

\newcommand{\MPL}[1]{Mod. Phys. Lett.\ {\bf #1}\ }

\newcommand{\AP}[1]{Ann.  Phys.\ {\bf #1}\ }

\def\lsim{\;\raise0.3ex\hbox{$<$\kern-0.75em\raise-1.1ex\hbox{$\sim$}}\;}
\def\gsim{\;\raise0.3ex\hbox{$>$\kern-0.75em\raise-1.1ex\hbox{$\sim$}}\;}

\newcommand{\eqref}[1]{(\ref{#1})}
\newcommand{\CR}{\nonumber \\}
\newcommand{\ie}{{\it i.e. }}
\newcommand{\eg}{{\it e.g. }}

\begin{document}
\begin{titlepage}
\begin{flushright}
{\large \bf UCL-IPT-96-05}
\end{flushright}
\vskip 2cm
\begin{center}
\vskip .2in

{\Large \bf
A light stop and electroweak baryogenesis\footnote{Work supported in part by the EEC Science Project SC1-CT91-0729.}}
\vskip .4in

{\large D. Delepine\footnote{Research assistant of the National Fund for
the Scientific Research}, J.-M. G\'erard, R. Gonzalez Felipe and J. Weyers}\\
\vskip .15in
{\em Institut de Physique Th\'eorique}\\

{\em Universit\'e catholique de Louvain}\\
\vskip .15in

{\em     
B-1348 Louvain-la-Neuve, Belgium}

\end{center}

\vskip 2in

\begin{abstract}

The possibility of creating baryon asymmetry at the electroweak phase transition in the minimal supersymmetric standard model is considered for the case when  right-handed squarks are much lighter than  left-handed ones. It is shown that  the usual requirement $v(T_c)/T_c  \gsim 1$ for baryogenesis can be satisfied  in a range of the parameters of the model, consistent with present experimental bounds.   
\end{abstract}
\end{titlepage}

\newpage
\renewcommand{\thepage}{\arabic{page}}
\setcounter{page}{1}
\setcounter{footnote}{1}

The interesting idea that the baryon asymmetry of the Universe may take place at the electroweak phase transition has attracted a lot of attention in the past few years \cite{cohen}. The necessary requirements for such a process to occur, namely, baryon-number violation, $C$ and $CP$ violation and departure from thermal equilibrium, were first pointed out by Sakharov \cite{sakharov}. Although these three criteria are in principle satisfied in the standard model, it is unlikely that the asymmetry can be generated within its framework at the electroweak phase transition. Indeed, the effects coming from the $CP$-violating phase in the Cabbibo-Kobayashi-Maskawa matrix seem to be  too small to explain the observed baryon number asymmetry \cite{gavela}. Furthermore, to produce this asymmetry it is necessary that processes which could wash out the latter asymmetry at the electroweak phase transition be out of equilibrium. In other words, the interaction rate for such processes should be slower than the expansion rate of the Universe. It has been shown \cite{shapo} that when  the electroweak phase transition is of the first-order,  a criterion for the latter condition to be satisfied is 
$$
\frac{M_{\rm sph}(T_c)}{T_c} \gsim 45\ ,
$$
where $M_{\rm sph}(T_c) = \frac{4 \pi}{g} v(T_c)  B(\lambda/g^2)$ is the sphaleron mass, $B$ is a constant which in the standard model ranges between $1.5 \le B \le 2.7$ for $0 \le \lambda/g^2 < \infty$ and $T_c$ is the critical temperature for the phase transition. Thus, suppression of the anomalous baryon number violation  after the transition requires a large jump in the Higgs vacuum expectation value $v(T)$ during the transition \cite{shapo}: 

\be
\frac{v(T_c)}{T_c} \gsim 1\ .
\label{1}
\ee

It is precisely this requirement \eqref{1} which imposes a severe constraint on the standard model, since it implies an upper bound on the Higgs mass, placed by perturbative \cite{carrington} and non-perturbative \cite{kajantie} calculations  in the experimentally ruled out region $m_H < 60$~GeV. 
 Therefore, it is of interest to investigate the possibility for the baryon asymmetry of the Universe to be generated at the electroweak scale in extensions of the standard model and in particular, in the minimal supersymmetric standard model (MSSM), which not only predicts a light Higgs boson but also contains new phases as sources of $CP$-violation \cite{gerard}. Previous numerical studies of the finite-temperature effective potential of the MSSM  have shown that the strength of the first-order electroweak phase transition is slightly enhanced in the presence of light stops \cite{giudice, espinosa}.  

In this letter we want  to show that  in  a simplified scenario described by the MSSM and for  a range of  the parameters of the model consistent with the present experimental bounds, there is still room for the constraint  \eqref{1} coming from baryogenesis to be satisfied. To make our presentation  more transparent, we shall keep only the  dominant  zero- and finite-temperature contributions to the effective potential. As a result our analysis can be given in an analytical form.  

The starting point for our discussion will be the one-loop effective potential. Let us recall that in the MSSM there are two complex scalar doublets, which after the spontaneous symmetry breaking give rise to five physical Higgs bosons: two charged, two $CP$-even and one $CP$-odd scalars. We shall assume that in the low-energy theory only one linear combination of the  $CP$-even neutral scalars remains light, while all the other Higgs bosons and supersymmetric partners of the usual standard model particles are heavy with  a mass of the order of the global supersymmetry breaking scale. In this case,  the tree level scalar potential for the real component of the lightest Higgs boson reads

\be
V_{\rm tree}=\frac{1}{2}\mu^2h^2+\frac{1}{32}\tilde{g}^2\cos^2 2\beta \ h^4\ ,
\label{2}
\ee 

\vspace{1 cm}
\noindent
where $h=h_1\cos \beta +h_2\sin \beta$, $\tilde{g}^2=(g^2_1+g^2_2)$ and $g_{1,2}$ are the $U(1)$ and $SU(2)$ gauge couplings respectively.

The main contribution to the one-loop effective potential comes from the  radiative corrections due to top and stop loops and it can be written in the form \cite{chaichian} 

\be
V_1=\frac{N_c}{16\pi^2}\int\limits^{\Lambda^2}_0 dk^2\,k^2 \left\{ \ln 
\left(1+\frac{m^2}{k^2+m_R^2}\right)  +  \ln \left(1+\frac{m^2}{k^2+m_L^2}\right) -
2  \ln \left(1+\frac{m^2}{k^2}\right) \right\} \ ,
\label{3}
\ee
where
\be
m \equiv \frac{g_t h}{ \sqrt{2}}
\label{4}
\ee
 is the field-dependent top quark mass, $g_t=h_t \sin \beta$  is the top Yukawa coupling to the scalar $h$, $N_c=3$ is the number of colours and  $\Lambda$ is an ultraviolet cutoff  (which does not play an important role in our discussion).  The parameters $m_L$ and $m_R$ are the soft supersymmetry-breaking mass parameters of the  left- and right-handed stops respectively. For simplicity, we have neglected  in Eq.\eqref{3}  the $D$-term contributions to the stop squared masses and possible left-right stop mixing effects. (A brief comment on the latter will be given in the conclusion). In other words, the field-dependent stop masses are given in our approximation by

\be
m^2_{\tilde{t}_{L,R}} = m^2_{L,R}+ m^2\ .
\label{5}
\ee

Assuming $m, m_L, m_R \ll \Lambda$ and neglecting terms that vanish as $\Lambda \rightarrow \infty$, the zero-temperature one-loop effective potential $V_0=V_{\rm tree}+V_1$  reads 

\bea
V_0 & =&  \frac{1}{2}\mu^2h^2+\frac{1}{32}\tilde{g}^2\cos^2 2\beta \ h^4 -
\frac{3}{8 \pi^2}\left\{ \frac{m^4}{2} \ln m^2 -  \frac{(m^2+m_L^2)^2}{4} \ln (m^2+m_L^2) \right. \CR
& -&  \left. \frac{(m^2+m_R^2)^2}{4} \ln (m^2+m_R^2) + \frac{m^2}{4} (m_L^2+m_R^2)  (2 \ln \Lambda^2 +1) \right\}\ .
\label{6}
\eea

The parameter $\mu$ in Eq.\eqref{6} can be expressed in terms of the physical masses  after the minimization of  the potential $V_0$. For the Higgs mass $m^2_h=(\partial^2V_0/\partial h^2)|_{h=v_0}$ one finally obtains

\be
m^2_h =  m^2_Z\cos^2 2\beta +\frac{3}{4\pi^2}\;\frac{m^4_t}{v_0^2}\ln \left[
\left(1+\frac{m^2_L}{m^2_t}\right) \left(1+\frac{m^2_R}{m^2_t}\right) \right]  \ ,
\label{7}
\ee
where $m_Z=\tilde{g}v_0 / 2$ is the $Z$-boson mass and $m_t=g_tv_0/\sqrt{2}$ is the  top quark mass, $v_0=246\ {\rm GeV}$. 

Let us now consider the finite-temperature contributions to the effective potential \eqref{6}. We shall proceed in a standard fashion by making use of the high- or low-temperature expansions \cite{dolan} of the effective  potential depending on the masses of the particles involved. At this point it is worth recalling that in the case of bosons with vanishing masses in the symmetric phase, the high-T expansion contains a cubic (in the field) term, which implies the coexistence of two minima at the critical temperature and thus, the corresponding phase transition is of first-order. This is precisely the case in the standard model due to the presence of  $W$ and $Z$ gauge bosons \cite{linde}. However, in the latter case the transition turns out to be only weakly first-order. 

 On the other hand,  in the MSSM the strength of the phase transition can be enhanced by the contribution of the $\tilde{t}_{L,R}$ stop fields. Since the masses of the stops do not vanish at $h=0$ (cf. Eq.\eqref{5}),  they do not induce a cubic term in the high-$T$ expansion of the  effective potential. A possible scenario for making stronger  the electroweak phase transition in the MSSM would be then the one where at least one of the soft supersymmetry-breaking mass parameters $m_{L,R}$ is light enough.  The latter scenario is naturally implemented, \eg, in the framework of  minimal supergravity \cite{nilles}, where $m_R=m_L$ at the grand unification scale (universality of the soft masses),  but $m_R$ can be much lighter than $m_L$ at the low-energy scale. Indeed, if the large top quark mass triggers the breakdown of $SU(2) \times U(1)$ through radiative corrections,  there is a well-known 3:2:1 hierarchy \cite{nilles} in the renormalization group equations for the Higgs scalar $h_2$, right-handed stop $\tilde{t}_R$ and left-handed stop $\tilde{t}_L$ squared masses,  respectively. This hierarchy would naturally ensure the color and electric charge conservation ($m_2^2 < 0$) as well as baryogenesis ($m_R^2 < m_L^2$). 

In what follows we shall assume that  $m_R  \ll m_L$  at the electroweak scale, \ie that the light stop is mainly right-handed, and also that $m_L \gg m_t$, so that $m_L$ characterizes the supersymmetry breaking scale. Under these assumptions, the left-handed stop is relevant only to the effective potential at zero temperature, since  its finite-$T$ contribution  is Boltzmann suppressed (low temperature expansion). On the contrary, the right-handed stop  gives an important contribution to the finite-temperature  effective potential and will be responsible for the  enhancement of the strength of the first-order phase transition. Keeping  the relevant terms for the gauge bosons, top quark and right-handed stop  in the high temperature limit \cite{dolan}, finally we obtain the temperature contribution to the effective potential

\bea
V_T&=& \frac{m^2 T^2}{2}+\frac{3 m^4}{16 \pi^2}\left( \ln \frac{m^2}{T^2} + C_F\right) 
       - \frac{2m_W^3+m_Z^3}{6\pi v_0^3} T h^3 \CR
      & -& \frac{(m_R^2+m^2)^{3/2} T}{2\pi}
       - \frac{3 (m_R^2+m^2)^2}{32 \pi^2} \left( \ln \frac{m_R^2+m^2}{T^2} -C_B\right)\ ,
\label{8}
\eea
where $C_F=-2.64$ and $C_B=5.41$.

Adding Eqs.\eqref{6} and \eqref{8}  and assuming $m_L \gg m_t$, we can write the finite-$T$ one-loop effective potential in the compact form

\be
V(h,T)=M^2(T) h^2-\delta(T) h^3 - a(T) (h^2+b^2)^{3/2} + \lambda(T) h^4\  ,
\label{9}
\ee
where 
\bea
M^2(T)= &-&\frac{1}{4}m^2_Z\cos^2 2\beta -\frac{3}{16\pi^2}\;\frac{m^2_t}{v_0^2}\left\{ m_t^2 \ln \left[\left(1+\frac{m^2_L}{m^2_t}\right) \left(1+\frac{m^2_R}{m^2_t}\right) \right] \right. \CR
           &+& \left. m_L^2 \ln \left(1+\frac{m_t^2}{m_L^2}\right) + m_R^2 \ln \left(1+ \frac{m_t^2}{m_R^2}\right) \right\} \CR
    &+& \frac{m_t^2}{2 v_0^2} T^2 + \frac{3}{16 \pi^2} \frac{m_t^2}{v_0^2} m_R^2 \left( 2\ln \frac{T}{m_R} + C_B-\frac{1}{2}\right)\ ,
\label{10}
\eea
\be
\delta(T)=\frac{2 m_W^3+m_Z^3}{6\pi v_0^3} T \ ,
\label{11}
\ee
\be
a(T)=\frac{m_t^3 T}{2\pi v_0^3}\ , \;\; b=\frac{m_R v_0}{m_t}\ ,
\label{12}
\ee
\be
\lambda(T)=\frac{1}{8} \frac{m_Z^2}{v_0^2} \cos^2 2\beta - \frac{3}{16\pi^2} \frac{m_t^4}{v_0^4}\left( \ln \frac{T}{m_L} - C_F - \frac{1}{2} C_B -\frac{3}{4}\right)\ .
\label{13}
\ee

Next we can obtain an improved one-loop finite temperature effective potential by  including  the  resummation of  the leading infrared-dominated  higher-loop contributions, related to the so-called daisy diagrams \cite{dolan}. In our present case, this will amount to replacing  the constant $b$  in Eqs.\eqref{9} and \eqref{12} by the temperature-dependent function

\be
b(T)= \frac{v_0}{m_t} \sqrt{m^2_R+\Pi_R (T)} \ ,
\label{14}
\ee
where \cite{espinosa}
\be
\Pi_R (T) = \frac{4}{9} g^2_3 T^2 + \frac{m_t^2}{3 v_0^2}\left( 2 + \tan^{-2} \beta \right) T^2
\label{15}
\ee 
is the finite temperature self-energy contribution to the right-handed stop and $g_3$ is the strong gauge coupling. Note that such a replacement allows us  to take also into consideration  the region  $m_R^2 < 0$, provided  that the shifted stop mass $m^2_{\tilde{t}_R} = m^2_R+ \Pi_R(T) + m^2$ remains positive and a colour-breaking global minimum is absent at $T=0$.

To establish  the dependence of  the vacuum expectation value  $v(T)$ of the Higgs scalar $h$ on the temperature, we should solve the equation for the extrema arising from the minimization of  the potential \eqref{9}, \ie the equation

\be
2 M^2(T)-3 \delta(T) v(T) -3 a(T) (v^2(T)+b^2(T))^{1/2}+4 \lambda(T) v^2(T)=0\ .
\label{16}
\ee

We are  interested in the solution of the equation \eqref{16} at  two specific critical temperatures, namely the temperature $T_0$ at which the potential \eqref{9} has a symmetry-breaking minimum and  is flat at the origin (end of the phase transition), 
\be
\left. \frac{\partial^2 V}{\partial h^2} \right|_{h=0, T=T_0} = 0\ ,
\label{17}
\ee
and the temperature $T_1$ at which both, the symmetric and symmetry-breaking, minima are degenerate (onset of the phase transition), 
\be
V(0,T_1)=V(v(T_1), T_1)\  .
\label{18}
\ee 
The phase transition will then take place at some temperature $T_c$, $T_0 \lsim T_c \lsim T_1$ and the following relations hold:
\be
\frac{v(T_1)}{T_1} \lsim \frac{v(T_c)}{T_c} \lsim \frac{v(T_0)}{T_0}\ .
\label{19}
\ee
   
Let us first find the critical temperature $T_0$. From Eqs.\eqref{9} and \eqref{17} it follows that

\be
 M^2(T_0) - 3 a(T_0) b(T_0) = 0\  ,
\label{20}
\ee
which with the use of Eqs.\eqref{10}, \eqref{12} and \eqref{14}  is equivalent to the equation

\bea
T^2_0 &-&\frac{3 T_0}{2\pi}\sqrt{m^2_R+\Pi_R(T_0)}+\frac{3 m_R^2}{4\pi^2} \ln 
\frac{T_0}{m_R} = \frac{m^2_Z v_0^2}{2m_t^2} \cos^2 2\beta +\frac{3}{8\pi^2}\left\{ \left(\frac{1}{2}- C_B\right) m_R^2 \right. \CR
 &+& m_t^2 \ln \left[\left(1+\frac{m^2_L}{m^2_t}\right) \left(1+\frac{m^2_R}{m^2_t}\right) \right]   + \left. m_L^2 \ln \left(1+\frac{m_t^2}{m_L^2}\right) + m_R^2 \ln \left( 1+ \frac{m_t^2}{m_R^2} \right) \right\}\  . \CR
\left. \right.
\label{21}
\eea

Substituting $M^2(T_0)$ from relation \eqref{20} into  Eq.\eqref{16}, we arrive at the cubic equation:

\be
\left. v^3-\frac{3\delta}{2\lambda} v^2 + \frac{9\delta^2+24ab\lambda-9a^2}{16\lambda^2} v -\frac{9ab\delta}{8\lambda^2} \right| _{T=T_0}= 0\ .
\label{22}
\ee

From  Eqs.\eqref{21} and  \eqref{22} we determine the ratio $v(T_0)/T_0$. As an illustrative example,  we have plotted  in Fig.1  (solid line)  the latter ratio as a function of the lightest stop mass $m_{\tilde{t}_R} = \sqrt{m_R^2+m_t^2}$ for the specific values of  $\tan \beta = 1.5$,  $m_L=1$~TeV  and $m_t=174$~GeV. We notice that  in  the  range of values  $150 \lsim  m_{\tilde{t}_R} \lsim 185$~GeV  the  ratio $v(T_0)/T_0  \gsim 1$. 

Next we consider the critical temperature $T_1$. In this case we have the system of coupled equations defined by \eqref{16} and \eqref{18}, which are easily solved numerically. The curve for the ratio $v(T_1)/T_1$ as a function of the stop mass  $m_{\tilde{t}_R}$ is also presented  in Fig.1 (bold-dashed line) , for the same values of $\tan \beta, m_L$ and $m_t$ given above. We see that at any temperature the latter ratio is, as expected, smaller than the corresponding one $v(T_0)/T_0$, but  the constraint $v(T_1)/T_1  \gsim 1$ is nevertheless satisfied in the region $150 \lsim  m_{\tilde{t}_R}  \lsim 168$~GeV.  

We remark that as $m_L$ decreases and for a fixed value of $\tan \beta$, the ratios $v(T_0)/T_0$ and $v(T_1)/T_1$ will increase, thus implying a wider range of values for $m_{\tilde{t}_R}$ where the constraint \eqref{1} is satisfied. The Higgs mass (cf.  Eq.\eqref{7}), in turn, will decrease as $m_L$ decreases, but  for $500\;  \rm{GeV} \lsim m_L \lsim 1\;  \rm{TeV}$ and $\tan \beta = 1.5$ we have $60 \lsim m_h \lsim 75$~GeV, consistent with the present experimental bounds.  

The dependence of the ratio $v(T)/T$ on $\tan \beta$ is plotted in Fig.2 for the particular values of $m_R=0$ and $m_L=500$~GeV. We see that as $\tan \beta$ increases the latter ratio decreases and for $\tan \beta = 1$ it reaches its maximum value. This fact  justifies our choice of values of  $\tan \beta \simeq 1$. Notice however that  $\tan \beta = 1$ implies a low value for  the Higgs mass defined  in Eq.\eqref{7}.  

Finally, we shall briefly comment on the possible mixing in the stop sector. Its inclusion in our analysis would  tend to weaken the first-order phase transition mainly due to the fact that it reduces the third term in the finite-temperature effective potential \eqref{9}. Indeed, assuming that $A_{\tilde{t}}$ is the effective stop mixing mass parameter, then the lightest stop mass will be approximately given by  $m^2_{\tilde{t}} = m_R^2 + m^2 (1-A^2_{\tilde{t}} / m_L^2)$ instead of the expression in \eqref{5}. 
 
To conclude,  under the assumption that the soft supersymmetry breaking parameter for the right-handed stop  is much smaller than the left-handed one ($m_R \ll m_L$), we have shown that  the baryogenesis constraint  \eqref{1} can be satisfied within the framework of the MSSM. This conclusion holds  for a range of values of the parameters  consistent with the present experimental results. In particular, our results remain valid for a right-handed stop  with a mass close to  the value of  the top quark mass and  for values of  $\tan \beta$ close to 1. 

{\it Note added:} While this work was in preparation we received the paper \cite{carena} where the authors perform a detailed numerical analysis of the effective potential  of the MSSM  and show that for low values of $\tan \beta$, $m_{\tilde{t}} \lsim m_t$ and  $m_h \lsim m_Z$ the ratio  $v(T_1)/T_1$ can indeed become significantly larger than one. We have checked that our results obtained using  the simplified  effective potential \eqref{9} are in agreement with those of Ref.\cite{carena} in the 10 - 15~\% range.  From this  we can conclude that  the terms  included in the improved one-loop effective potential given by  Eqs. \eqref{9}-\eqref{15} represent  the most important effects. Other terms  such as the additional gauge boson contributions at zero and finite temperature, the  D-terms as well as the finite-temperature effects of the left-handed stop are negligible in the range of the parameters of the MSSM we are dealing with.

%\newpage
\input{epsf.sty}
\begin{figure}[t]
\leavevmode
\begin{center}
\mbox{\epsfxsize=15.cm\epsfysize=10.cm\epsffile{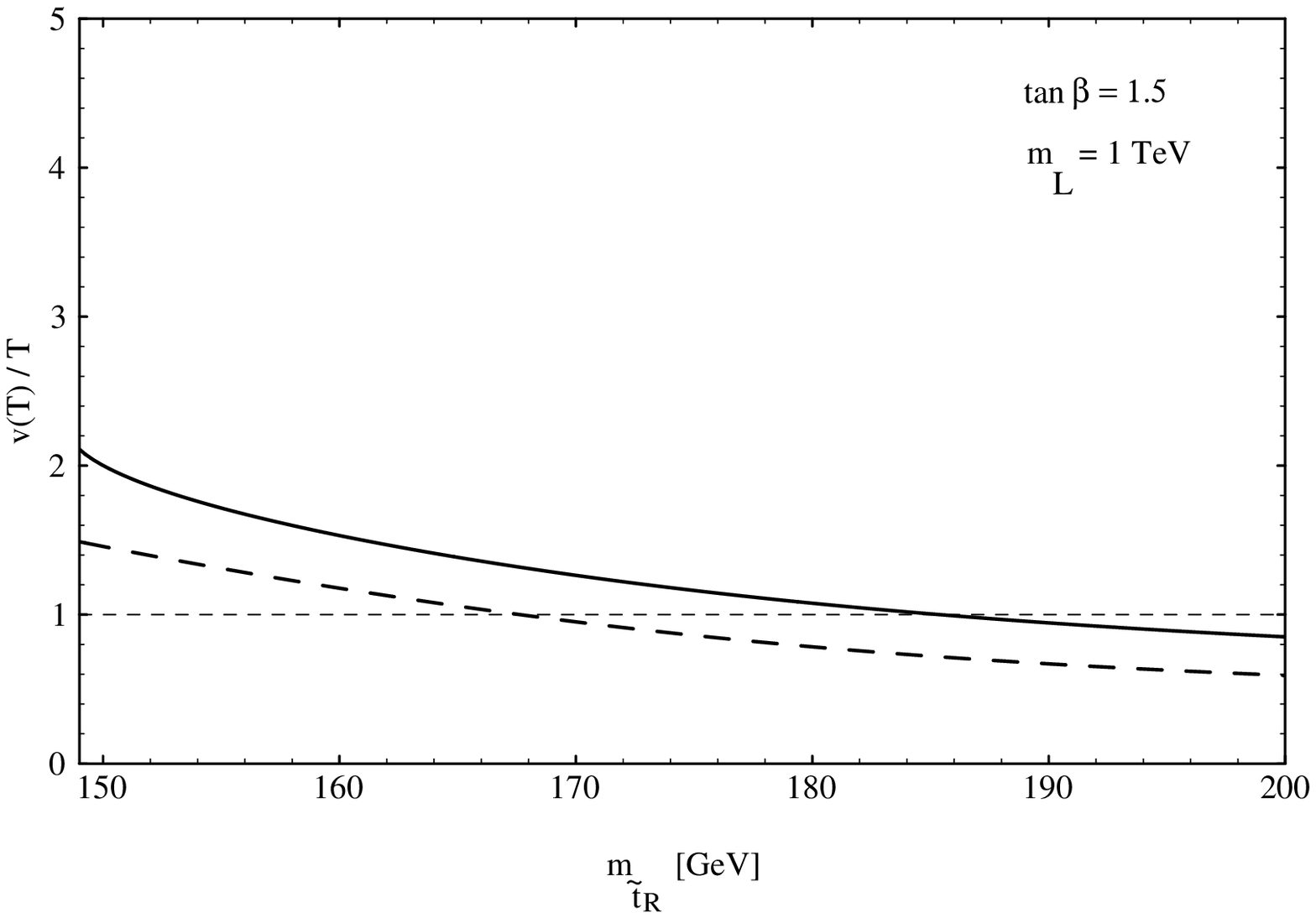}}
\end{center}
\caption{ Dependence of the ratios $v(T_0)/T_0$ (solid line) and $v(T_1)/T_1$ (bold-dashed line)  on the right-handed stop mass $m_{\tilde{t}_R}$ for the particular values of $\tan \beta = 1.5$, $m_L = 1$~TeV and $m_t=174$~GeV. }
\label{fig1}
\end{figure}

%\newpage
\input{epsf.sty}
\begin{figure}[t]
\leavevmode
\begin{center}
\mbox{\epsfxsize=15.cm\epsfysize=10.cm\epsffile{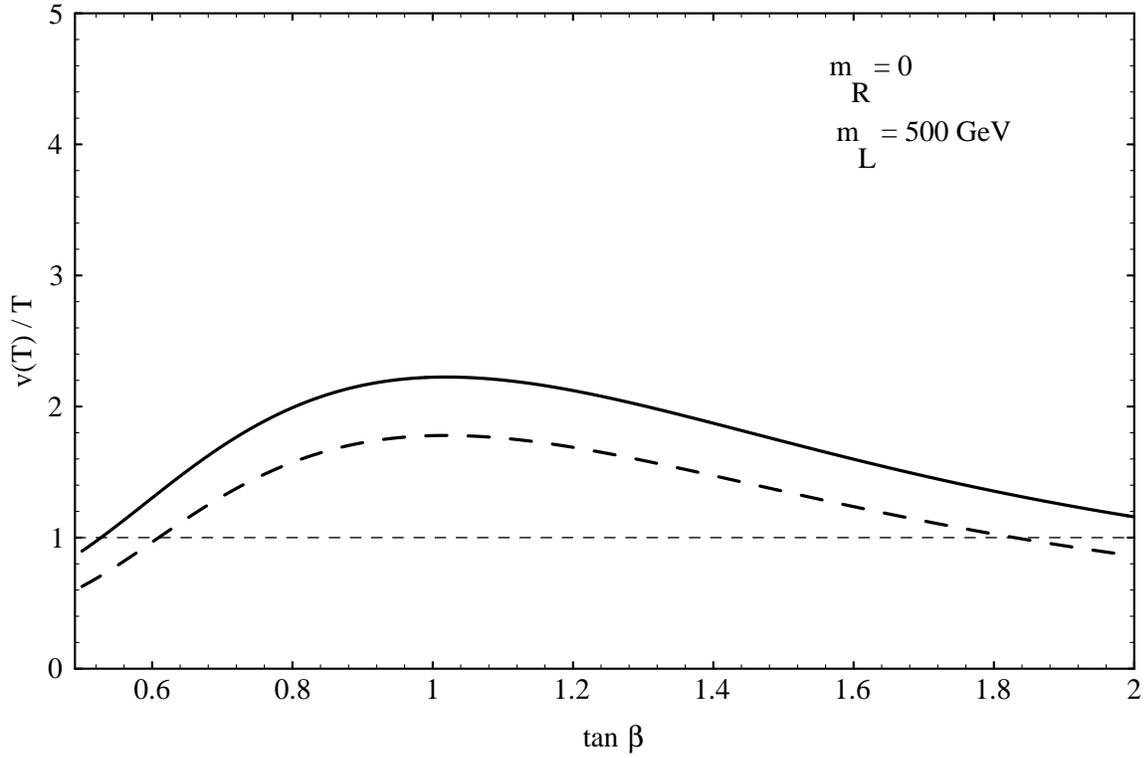}}
\end{center}
\caption{ The curves  $v(T_0)/T_0$ (solid line) and $v(T_1)/T_1$ (bold-dashed line)  as functions of $\tan \beta$ for $m_R=0$, $m_L=500$~GeV and $m_t=174$~GeV.}
\label{fig2}
\end{figure}

\end{document}